\documentclass[reprint,twocolumn,amsmath,amssymb,showpacs,floatfix,aps,prb]{revtex4}
\usepackage{color}
\usepackage{dcolumn}
\usepackage{bm}
\usepackage{graphicx}
\begin{document}
\preprint{APS/123-QED}
\title{Tunable rotons in square lattice antiferromagnets under strong magnetic fields}
\author{Yurika Kubo}
\email{kubo@kh.phys.waseda.ac.jp} 
\author{Susumu Kurihara}%
\altaffiliation[Also at ]{Department of Physics, Waseda University}
\affiliation{%
Department of Physics, Waseda University, 3-4-1 Okubo, Shinjuku, Tokyo 169-8555, Japan}%
\date{\today}
\begin{abstract}
Excitation spectra of square lattice Heisenberg antiferromagnets in magnetic fields are investigated by the spin-wave theory. It is pointed out that a rotonlike structure appears in a narrow range of magnetic fields, as a result of strong nonlinear effects. It is shown that the energy gap and the mass of the ``roton'' are quite sensitive to the magnetic field: the roton gap softens rapidly and eventually closes as a precursor of a quantum phase transition. The possibility of the experimental observation of the roton and a new ground state after its softening are discussed.
\end{abstract}
\pacs{75.10.Jm,75.30.Ds,75.50.Ee}
\maketitle
\section{INTRODUCTION}
Square lattice Heisenberg antiferromagnets (SLHAFs) are well known to have the N\'{e}el order at zero field, and their excitation spectra are well described by the linear spin wave (LSW) theory with small renormalization \cite{ZFSHAF}. However, an external magnetic field induces noncollinear structure resulting in nonlinear three-magnon interactions \cite{Mourigal,ZhitSCBApprox}. It should be stressed that the three-magnon interactions in SLHAFs can be controlled, from zero to sufficiently large values, simply by tuning external magnetic fields \cite{ZhitSCBApprox,Mourigal,ZhitRMP,EDResult}. These features are in sharp contrast to the triangular lattice antiferromagnets, which have a noncollinear 120$^\circ$ structure and therefore have strong nonlinear interactions even for zero field \cite{TriTheo,Zheng}. Accordingly, SLHAF is an ideal system to examine the effects of the three-magnon interactions. 

Theoretical calculations \cite{QMCResult,EDResult,ZhitSCBApprox,Mourigal} and experiments \cite{Masuda,Tsyrulin} on excitation spectra of SLHAF in fields show significant deviations from that of the LSW calculations. Frustration-induced noncollinear antiferromagnets also show such deviations \cite{TriTheo,Zheng,TriExp,TanaKagome}.

Zhitomirsky and Chernyshev \cite{ZhitSCBApprox} and Mourigal {\it et al}. \cite{Mourigal} proposed several methods to calculate the magnon spectra in high fields, where the three-magnon interactions are strong \cite{ZhitSCBApprox,Mourigal,ZhitRMP,Quasi2DZhitGroup}. However, a self-consistent Born approximation (SCBA), which neglects vertex corrections, results in an unphysical gap in the acoustic mode \cite{ZhitSCBApprox}. This is due to a violation of the Ward-Takahashi identity, suggesting crucial importance of vertex corrections in such a renormalization. Mourigal {\it et al}. \cite{Mourigal} partially perform SCBA for $S\geq1$, still neglecting vertex corrections, and consider only the imaginary part of self-energy. They thus remove the unphysical gap \cite{Mourigal} but do not essentially solve the problem of the violation of the Ward-Takahashi identity. Last, Fuhrman {\it et al} \cite{Quasi2DZhitGroup} introduce an alternative idea of adding some interlayer interactions, rendering the system essentially three-dimensional. This, however, cannot be the solution to the difficulty in the purely two-dimensional model, and we still lack reliable results. 

We believe that the perturbation calculation is much more reliable than that of partial renormalization of self-energy since the former satisfies the Ward-Takahashi identity albeit in a trivial way. Thus, we calculate the nonlinear spin-wave spectra of the purely two-dimensional SLHAFs in fields within the simple second-order perturbation theory on the basis of the Zhitomirsky-Mourigal formalism \cite{ZhitSCBApprox,Mourigal}. 

Our calculation shows that a rotonlike minimum emerges in a quite narrow range of fields at about $3/4$ of the saturation field $H_{\rm s}$, a remarkable feature in the magnon spectrum which previous works \cite{ZhitSCBApprox,Mourigal,QMCResult,EDResult} might have overlooked. We also find that the roton gap drops steeply to zero as the field increases. 

This paper is composed as follows. First, we briefly introduce the spin-wave formalism following Zhitomirsky and Mourigal {\it et al}. Then, we show how the spin-wave spectra vary with fields within the second-order perturbation calculation. The main feature of the spectra is an appearance of a rotonlike minimum which responds sensitively to small changes of fields. Last, we discuss a new ground state after the softening of the roton and possibilities of confirming the roton feature. 

\section{MODEL}
In this section, the main points of the Zhitomirsky-Mourigal formalism on SLHAFs \cite{ZhitSCBApprox,Mourigal} are summarized. The Heisenberg Hamiltonian in a magnetic field $H$ is:
\begin{align}
\hat{\cal H}=J\sum_{<i,j>}{\bf S}_i\cdot{\bf S}_j- H\sum_i S_i^{z_0} \textrm{,} \label{hamiltonian}
\end{align} 
where $S_i^{\mu}\,(\mu\!=\!x_0,y_0,z_0)$ denote spin operators in the laboratory frame and $J$ denotes the nearest neighbor exchange constant. Then, we move from the laboratory frame to the rotating frame with spin operators $S_i^{\mu}\,(\mu\!=\!x,y,z)$:
\begin{align}
\begin{split}
S_i^{x_0} &= S_i^x \sin \theta+S_i^z {\rm e}^{{\rm i}{\bf Q}\cdot{\bf R}_i} \cos \theta \textrm{,}\hspace{7mm}
S_i^{y_0} = S_i^y \textrm{,} \\
S_i^{z_0} &= -S_i^x {\rm e}^{{\rm i}{\bf Q}\cdot{\bf R}_i} \cos \theta+S_i^z \sin \theta \textrm{,}
\end{split}
\end{align} 
where ${\bf Q}=(\pi,\pi)$ denotes the ordering wave vector. The canting angle $\theta$ is chosen to minimize the ground state energy: 
\begin{align}
\theta=&\sin^{-1} [h] \textrm{,}&h&=H/H_{\rm s}\textrm{,}
&H_{\rm s}&=8JS\textrm{.}\label{gscantangle}
\end{align}

We then perform the Holstein-Primakoff (HP) transformation:
\begin{align}
\begin{split}
S_i^{+}&=\sqrt{2S-a^{\dagger}_ia_i}\,a_i\textrm{,} \hspace{7mm} S_i^z=S-a^{\dagger}_ia_i\textrm{,}\\
S_i^{-}&=a^{\dagger}\sqrt{2S-a^{\dagger}_ia_i}\textrm{,}
\end{split}
\end{align}
where $a_i$ denotes HP bosons. We get
\begin{align}
\hat{\cal H}=\sum_{n=0}^{\infty}\hat{\cal H}_n \textrm{,}
\end{align}
where $\hat{\cal{H}}_n$ denotes the $n$ th-order term in HP boson operators \cite{Mourigal} and $\hat{\cal{H}}_1$ vanishes by determining $\theta$ correctly \cite{Mourigal,ZhitSCBApprox,Nikuni}. 

We perform the Fourier transformation and then the Bogoliubov transformation\cite{Mourigal,ZhitSCBApprox,Nikuni}:
\begin{align}
a_{\bf k}&=u_{\bf k}b_{\bf k}+v_{\bf k}b^{\dagger}_{-{\bf k}}\hspace{8mm}\left(\,u_{\bf k}^2-v_{\bf k}^2=1\,\right)\textrm{,}
\end{align}
where $b_{\bf k}$ denotes Bogoliubov bosons. $\hat{{\cal H}}_2$ is readily diagonalized, yielding an excitation spectrum:
\begin{align} 
\begin{split}
\epsilon_{\bf k} &= \sqrt{A_{\bf k}^2-B_{\bf k}^2} \textrm{,} \hspace{10mm}\gamma_{\bf k}=\frac{\cos k_x+\cos k_y}{2} \textrm{,} \\
A_{\bf k} &= 4JS(1+\gamma_{\bf k}\sin^2 \theta)\textrm{,} \hspace{2mm} B_{\bf k} = 4JS\gamma_{\bf k}\cos^2 \theta \textrm{.}\label{LSWenergy} 
\end{split}
\end{align}

We now focus on the three-magnon interaction Hamiltonian $\hat{\cal{H}}_3$:
\begin{align}
\begin{split}
\hat{\cal{H}}_3=&\frac{1}{2!}\sum_{{\bf k},{\bf q}}(b^{\dagger}_{\bf q}b^{\dagger}_{{\bf p}_1}b_{\bf k}+\textrm{H.c.})
\Phi_{1}(\bf{k},{\bf p}_1,\bf{q})\label{phi1}\\
&+\frac{1}{3!}\sum_{{\bf k},{\bf q}}(b^{\dagger}_{\bf k}b^{\dagger}_{{\bf p}_2}b^{\dagger}_{\bf{q}}+\textrm{H.c.})
\Phi_{2}(\bf{k},{\bf p}_2,\bf{q})\textrm{,}
\end{split}
\end{align}
where ${\bf p}_1\!=\!{\bf Q}\,+\,{\bf k}\,-\,{\bf q}$, ${\bf p}_2\!=\!{\bf Q}\,-\,{\bf k}\,-\,{\bf q}$, and $\Phi_{\alpha}(1,2,3)\propto\sin 2\theta$ ($\alpha\!=\!1\textrm{, }2$) are given in Refs. \onlinecite{Mourigal} and \onlinecite{ZhitSCBApprox} and they come into play only with noncollinear magnetic structures. Self-energy corrections, which are generated by Eq. (\ref{phi1}), are \cite{ZhitSCBApprox,Mourigal}
\begin{align} 
\Sigma^{(1)}({\bf k},\omega)&=\frac{1}{2}\sum_{\bf q}\frac{\left|\Phi_1({\bf k},{\bf p}_1,{\bf q})\right|^2}{\omega-\epsilon_{{\bf p}_1}-\epsilon_{\bf q}+i0}\textrm{,}\label{SIGMA1}\\
\Sigma^{(2)}({\bf k},\omega)&=-\frac{1}{2}\sum_{\bf q}\frac{\left|\Phi_2({\bf k},{\bf p}_2,{\bf q})\right|^2}{\omega+\epsilon_{{\bf p}_2}+\epsilon_{\bf q}-i0}\textrm{.}\label{SIGMA2}
\end{align}
Lowest order $1/S$ corrections generated by three-magnon couplings are given by following on-shell self-energy:
\begin{align} 
\Sigma_{\bf k}=\Sigma^{(1)}({\bf k},\epsilon_{\bf k})+\Sigma^{(2)}({\bf k},\epsilon_{\bf k})\textrm{.}\label{SIGWa}
\end{align}
We see from Eqs. (\ref{SIGMA1})-(\ref{SIGWa}) that the self-energy correction is especially large with a smaller energy difference between the initial and intermediate states. 

We also need to perform the Hartree-Fock decoupling in $\hat{\cal{H}}_4$, leading to a correction $\delta\epsilon^{\rm HF}_{\bf k}$, and take quantum corrections to the canting angle into account \cite{Nikuni}, yielding another correction $\delta\epsilon^{\theta}_{\bf k}$. Finally, we get the $1/S$ corrected spin-wave spectra \cite{ZhitSCBApprox,Mourigal},
\begin{align} 
\bar{\epsilon}_{\bf k}=\epsilon_{\bf k}+\Sigma_{\bf k}+\delta\epsilon^{\rm HF}_{\bf k}+\delta\epsilon^{\theta}_{\bf k}\textrm{.}
\end{align}

\section{MIXING OF THE ONE- AND TWO-MAGNON STATES}
The effects of the coupling between one- and two-magnon states on $\bar{\epsilon}_{\bf k}$ become quite strong at around $h\approx0.75$. There are two reasons for this. First, $\Phi_{1}(1,2,3)\!\propto\!\sin 2\theta$ \cite{ZhitSCBApprox,Mourigal}, which reflects the strength of hybridizations, takes the maximum value at around $h\approx0.75$. Second, the curvature on the acoustic mode becomes positive also for $h\gtrapprox0.75$ \cite{ZhitRMP,ZhitSCBApprox,Mourigal}. The relation between the positive curvature and the strong couplings is briefly discussed below based on previous works \cite{ZhitRMP,ZhitSCBApprox,Mourigal}. 

The curvature on the acoustic mode of the LSW spectrum increases monotonically as the field increases. The nonlinear interactions also increase since the higher curvature induces the stronger three-magnon interactions. 

The curvature changes its sign from negative to positive. This sign change first occurs at $h^{*}=2/\sqrt{7}\approx0.7559$ along the $\Gamma$-$M\!=\!(\pi,\pi)$ line near the $M$ point, as mentioned in Refs. \onlinecite{Mourigal} and \onlinecite{ZhitSCBApprox}, since the curvature is highest along the $\Gamma$-$M$ line. This sign change is important for discussing the spontaneous magnon decay because the positive curvature is required to satisfy the kinematic constraint:
\begin{align}
\epsilon_{\bf k}=\epsilon_{\bf q}+\epsilon_{{\bf Q}-{\bf q}+{\bf k}}\textrm{.}
\end{align}
The region where the energy conservation law holds spreads across the Brillouin zone with the spreading positive curvature as the field increases. The spreading decay region as a function of field is shown in Fig. 5 of Ref. \onlinecite{Mourigal}.

We note that an especially strong mixing of the one- and two-magnon states is expected near the threshold of the decay region, where the energy conservation law holds, since there are many processes which have infinitesimal energy differences between the initial and intermediate states [see Eq. (\ref{SIGMA1})]. Consequently, we focus on an intersection point of the decay threshold with the $\Gamma$-$M$ line, where the particularly strong hybridizations between the states are expected. 

\section{APPEARANCE AND SOFTENING OF ``ROTONS'' IN STRONG MAGNETIC FIELD}
Now, we discuss the magnon spectrum $\bar{\epsilon}_{\bf k}$ renormalized by three-magnon couplings given in Eq. (\ref{SIGWa}), corresponding to the $1/S$ corrections coming from the Holstein-Primakoff expansion of spin operators. 
\subsection{Appearance of a rotonlike minimum}
\begin{figure}[!t]
\begin{center}
\begin{minipage}{0.95\linewidth}
\begin{center}
\includegraphics[width=0.98\linewidth]{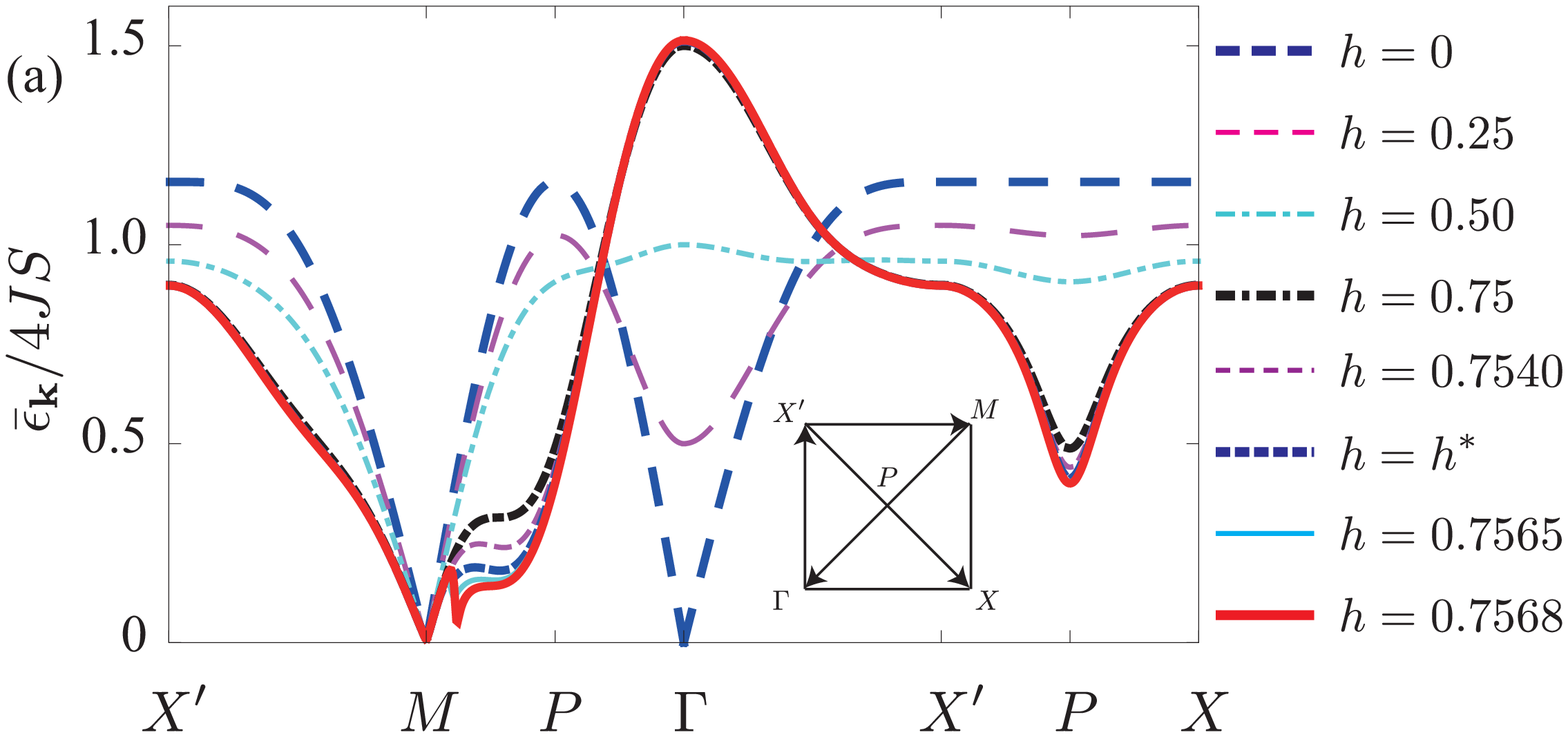}
\includegraphics[width=0.9\linewidth]{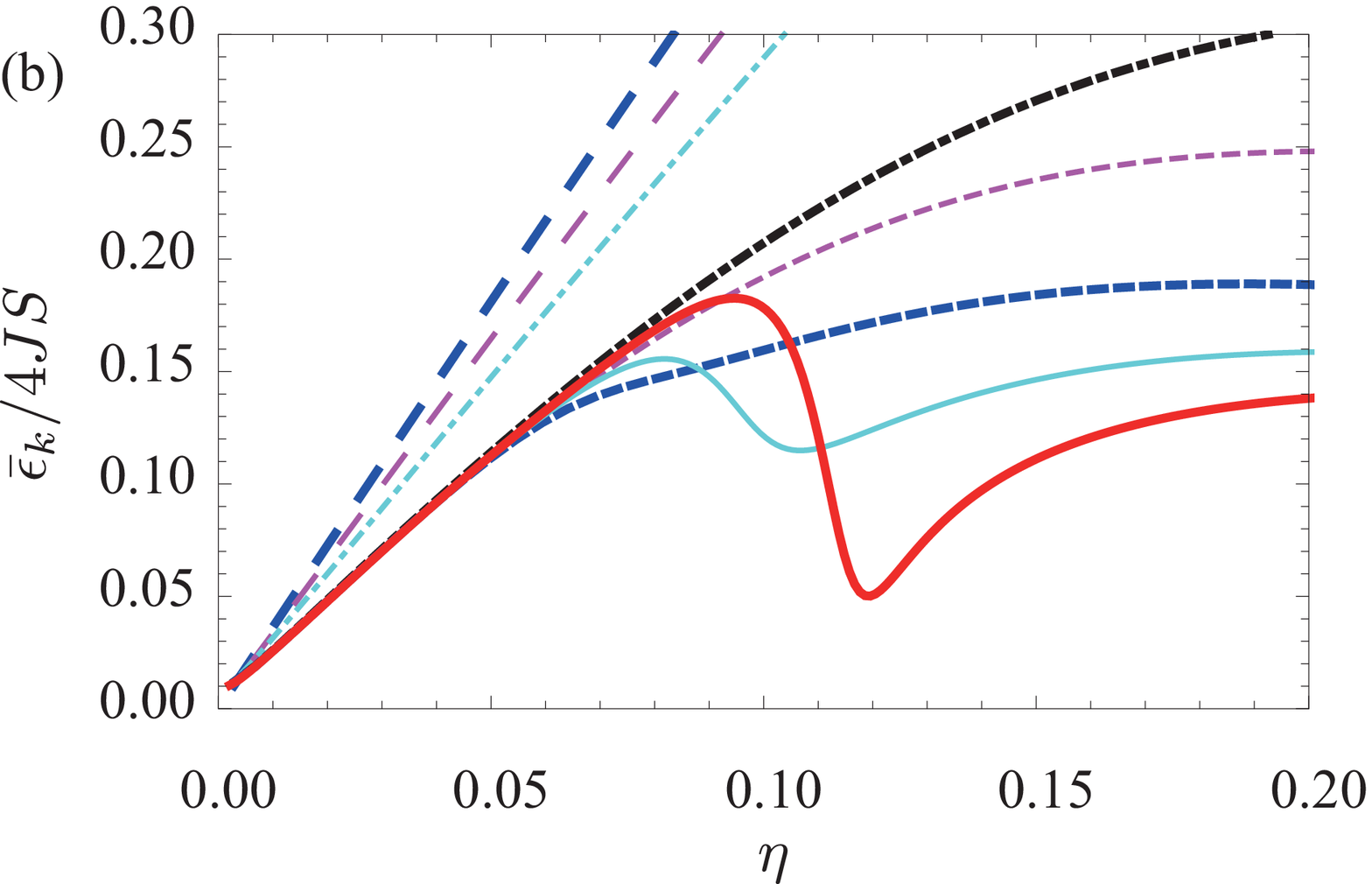}
\includegraphics[width=0.9\linewidth]{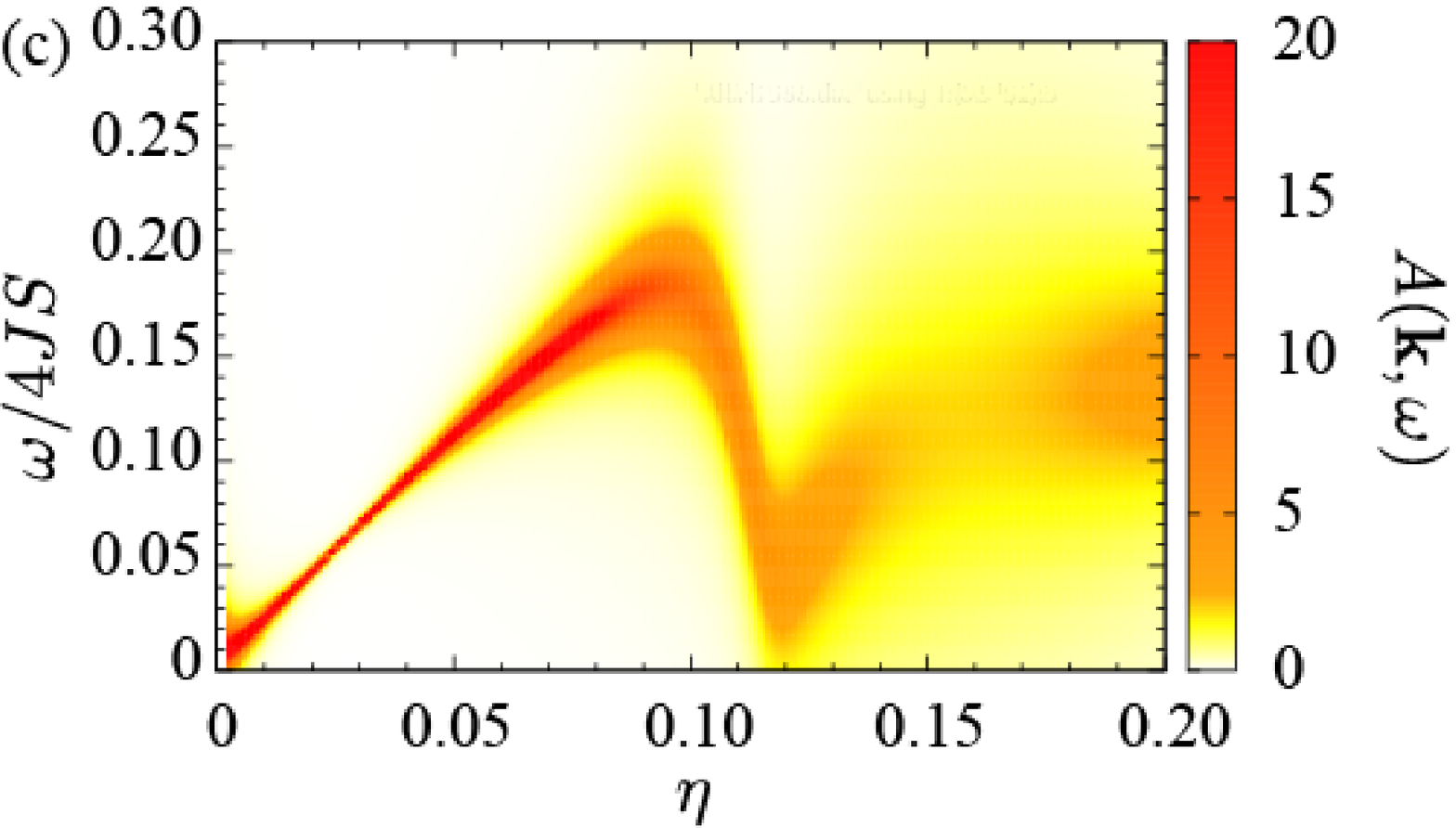}
\end{center}
\end{minipage}
\caption{(Color online) (a) Nonlinear spin-wave spectra $\bar{\epsilon}_{\bf k}$ for the $S=1/2$ system along the highly symmetric line calculated for several magnetic fields. The magnetic field $h$ corresponding to each line is shown on the right side. The spectrum along the $\Gamma$-$M$ line near the $M$ point varies drastically for small changes of $h$. There are apparent minima at $P$ point along the $X$-$X'$ line at finite fields. (b) Enlarged $\bar{\epsilon}_{\bf k}$ for $S=1/2$ SLHAF along the $\Gamma$-$M$ line near the $M$ point [$\pi(1\!-\!\eta,1\!-\!\eta), 0\leq\eta\leq0.20$].(c) Spectral weight $A ({\bf k}, \omega)$ (b) ($S=1/2$, $h=0.7568$). We notice a sizable broadening by magnon decay, but we still see the roton feature in the part with colored orange [$A({\bf k},\omega)\gtrsim 2$].}\label{GMXXmag}
\end{center}
\end{figure}

Figure \ref{GMXXmag}(a) shows the spectra for the $S=1/2$ SLHAF, calculated for several magnetic fields $h$. Rotonlike structure emerges near the $M$ point, which is enlarged in Fig. \ref{GMXXmag}(b) for clarity. Very sensitive responses to slight changes of $h$ are seen along the $\Gamma$-$M$ line, while little change is seen along the others. Therefore, we focus on $\bar{\epsilon}_{\bf k}$ along the $\Gamma$-$M$ line. 
\begin{figure}[!t]
\begin{center}
\begin{minipage}{0.9\linewidth}\centering
\includegraphics[width=0.9\linewidth]{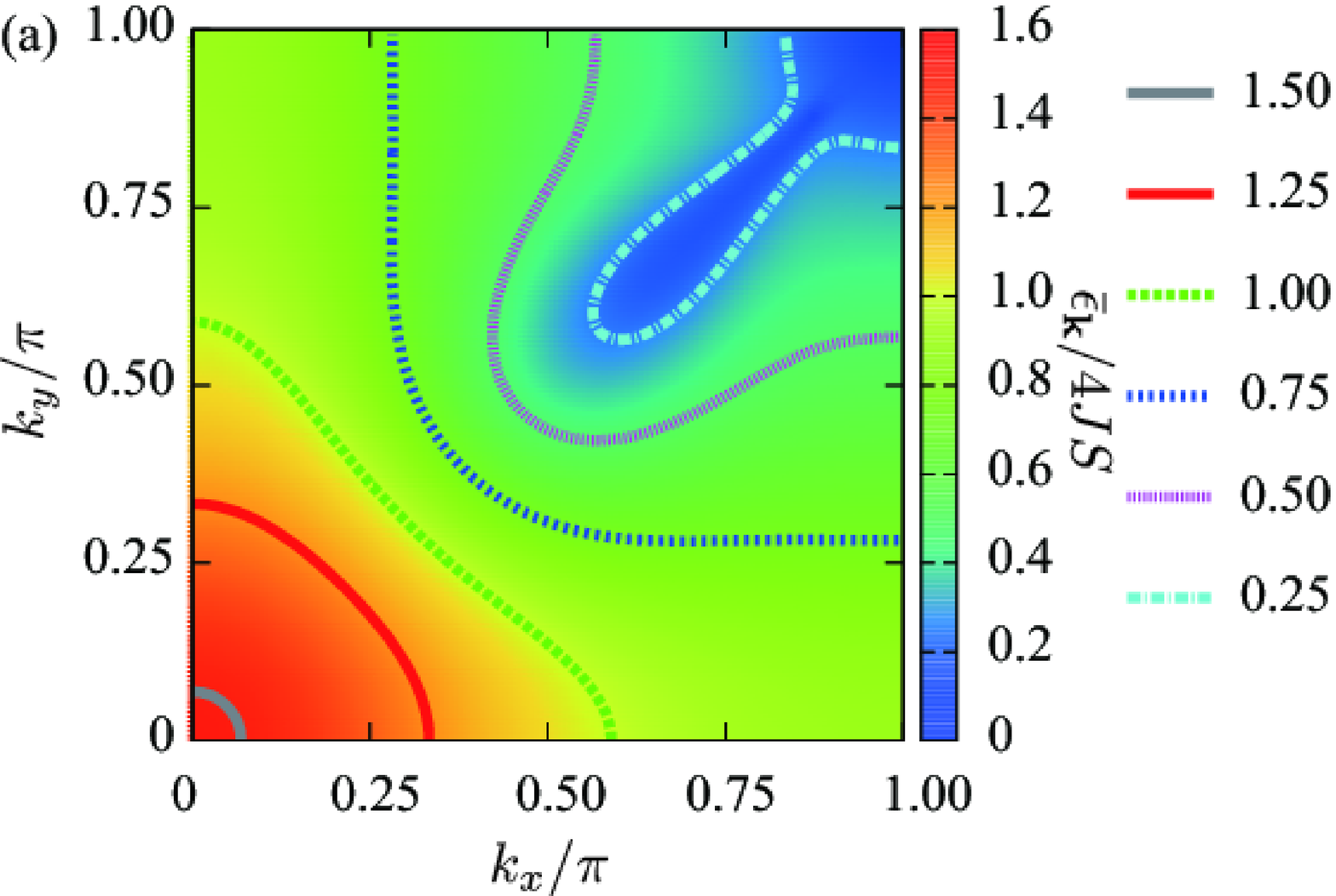}
\includegraphics[width=0.9\linewidth]{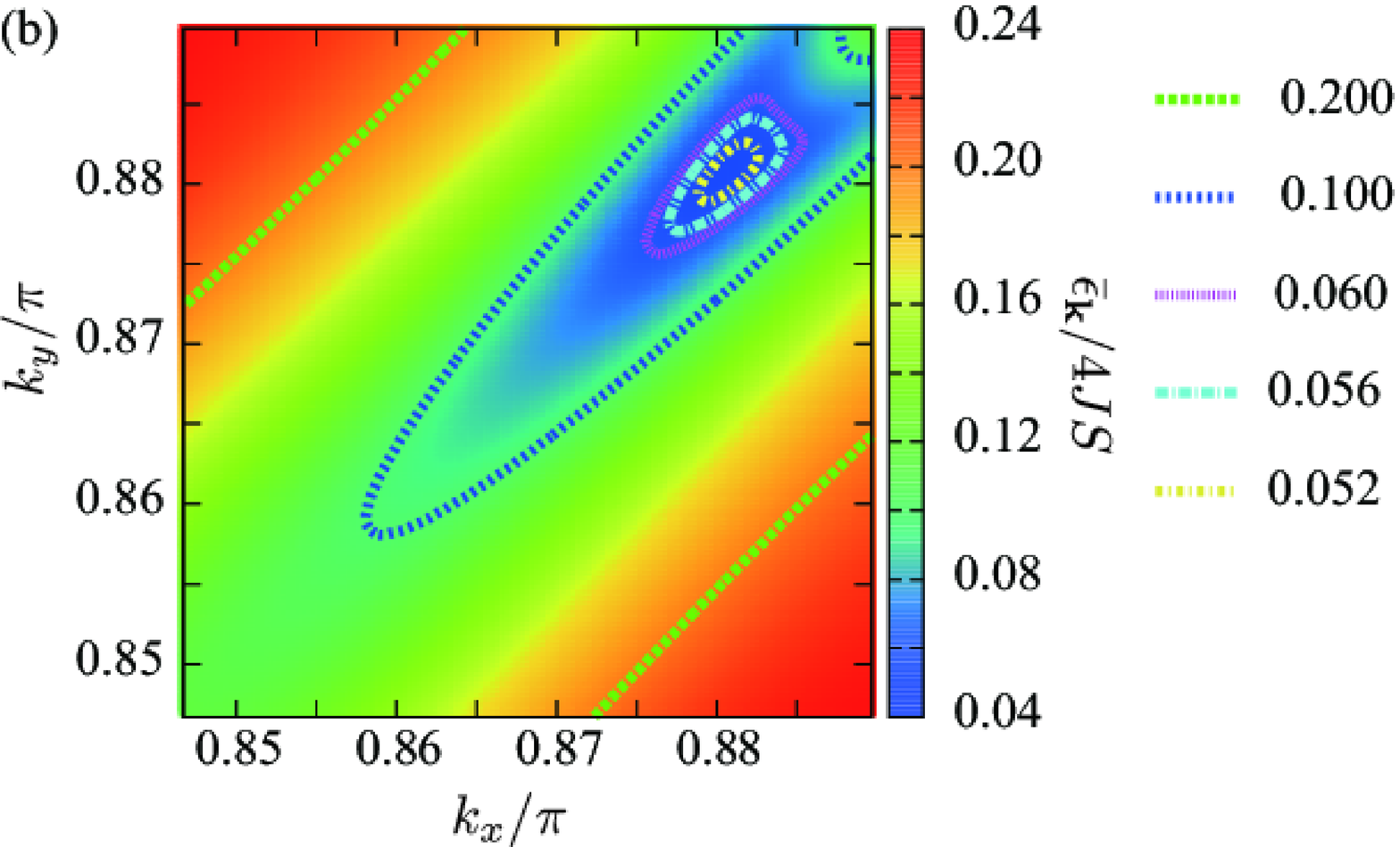}
\includegraphics[width=0.9\linewidth]{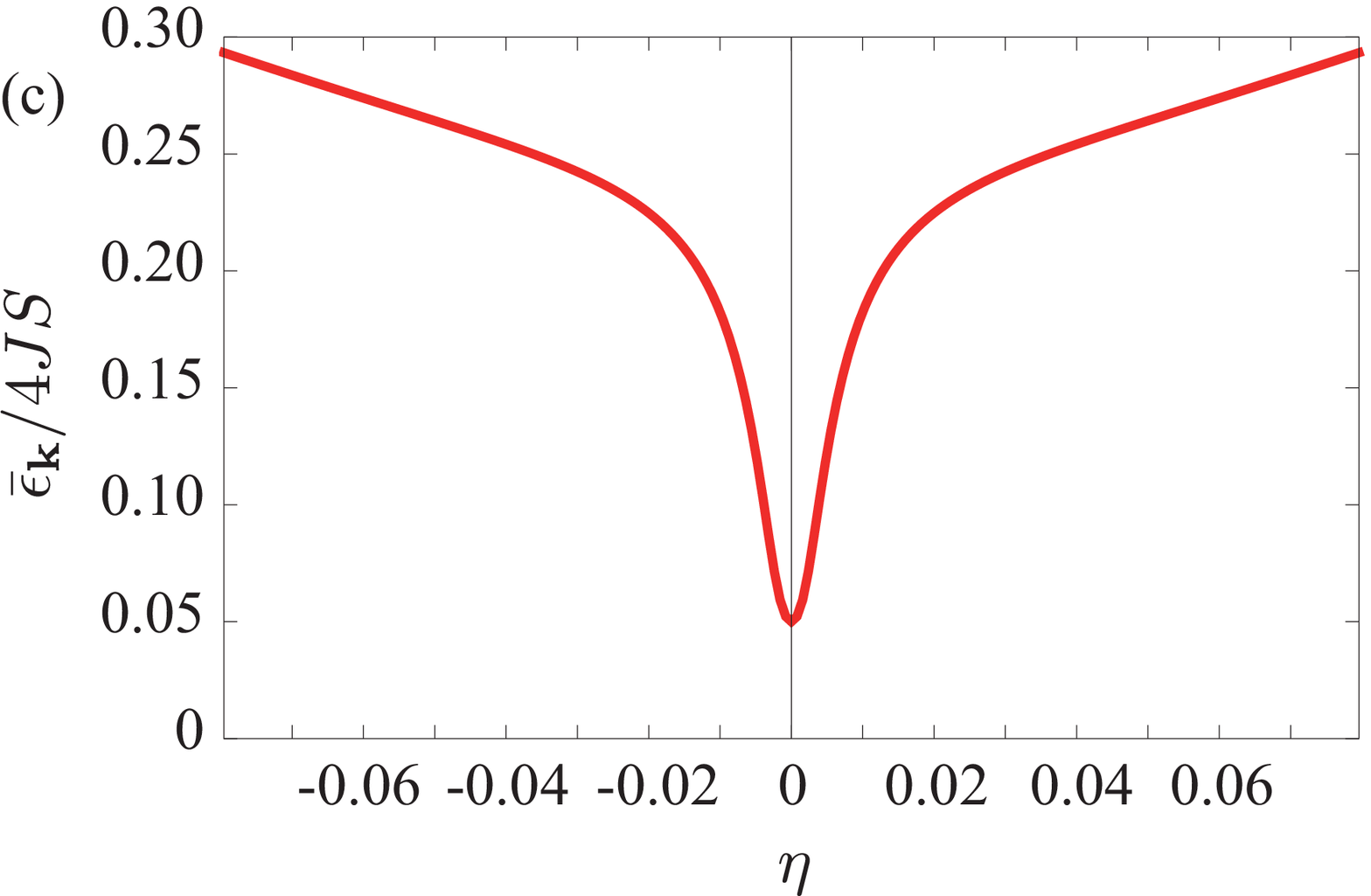}
\caption{(Color) (a) Contour plot of $\bar{\epsilon}_{\bf k}$ ($S\!=\!1/2$, $h\!=\!0.7568$) of the whole Brillouin zone. The energy corresponding to each contour is shown on the right side. (b) Enlarged contour plot of $\bar{\epsilon}_{\bf k}$ near the $M$ point. It is now clear that an energy minimum appears near the $M$ point. (c) Enlarged $\bar{\epsilon}_{\bf k}$ along the line $\pi(\eta_{\rm rot}\!-\!\eta,\eta_{\rm rot}\!+\!\eta)$, where $\eta_{\rm rot}$ denotes a roton wave vector $\pi \eta_{\rm rot}(1,1)$, perpendicular to the $\Gamma$-$M$ line. Note the sharpness of the minimum along this direction; the half width is on the order of 0.01. }\label{ContPlot}
\end{minipage}
\end{center}
\end{figure}
Enlarged spectra $\bar{\epsilon}_{\bf k}$ in Fig. \ref{GMXXmag}(b) remind us of the roton in the superfluid helium \cite{HeRoton}. We find that the wave vector, gap, and mass of a rotonlike structure change drastically as a result of a less than 1\% change in the magnetic field $h$ near $h^{*}=2/\sqrt{7}$. The gap gets smaller and smaller by increasing fields and finally vanishes, which indicates a quantum phase transition characterized by a certain modulation of the ground state. 

We show, in Fig. \ref{GMXXmag}(c), the spectral weight:
\begin{align}
A ({\bf k}, \omega) = \frac{1}{\pi}\frac{\zeta_{\bf k}}{(\omega-\bar{\epsilon}_{\bf k})^2+\zeta_{\bf k}^2}\textrm{,}
\end{align}
where $\zeta_{\bf k}$ denotes the imaginary part of the on-shell self-energy. We use red for an intensity stronger than the maximum value in the color bar. We now see that the roton feature in Fig. \ref{GMXXmag}(b) is not destroyed by magnon decay, despite the sizable broadening of the spectra.

The contour plot of $\bar{\epsilon}_{\bf k}$ ($S=1/2$, $h=0.7568$) of the whole Brillouin zone is shown in Fig. \ref{ContPlot}(a). We see that the strong three-magnon couplings at around $h\approx0.75$ induce anisotropic spectra along the $\Gamma$-$M$ line. We also show an enlarged $\bar{\epsilon}_{\bf k}$ ($S=1/2$, $h=0.7568$) near the $M$ point in Fig. \ref{ContPlot}(b). It is now clear that the minimum near the $M$ point is in fact a local minimum in the two-dimensional Brillouin zone, thus deserving the name roton. Figure \ref{ContPlot}(c) shows the roton spectrum near the roton wave vector along the line perpendicular to the $\Gamma$-$M$ line. We note that the minimum is much sharper along this line, with a half width of order 0.01. We discuss the origin of the sharpness in the next section. 
\begin{figure}[!t]
\begin{center}
\includegraphics[width=0.95\linewidth]{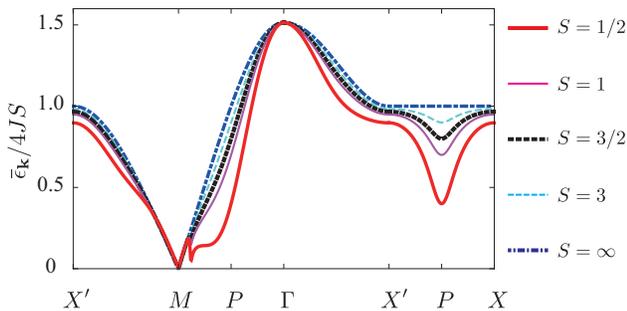}
\caption{(Color online) (a) Nonlinear spin wave spectra $\bar{\epsilon}_{\bf k}$ for the various $S$ including $S=\infty$ (LSW) at $h=0.7568$. The spin magnitude $S$ for each line is shown on the right side. Spectra become more classical for larger $S$, but no qualitative changes are observed. It is also suggested that the quantum effects are stronger along the $\Gamma$-$M$ line than the others.}
\label{Sgeq05}
\end{center}
\end{figure}
We also observe apparent minima at point $P$ [$(\pi/2,\pi/2)$] along the $X$-$X'$ line. However, $P$ point is not really a local minimum in the Brillouin zone since it is on a downhill slope along the $\Gamma$-$M$ line.

Rotonlike minima near the $M$ point have not been reported, to our knowledge. This might be due to the extreme narrowness of the field range where the roton can exist. Concerning the $P$ point features, recently synthesized, almost ideal SLHAF $\mathrm{Cu(pz)_2(ClO_4)_2}$ [$pz$ stands for the pyrazine molecule] exhibits stronger response to fields on the $P$ point than that of the $X'$ point \cite{Tsyrulin}. This is qualitatively consistent with our results in Fig. \ref{GMXXmag}(a) and Refs. \onlinecite{Mourigal} and \onlinecite{EDResult}. 

The dependence of the magnon spectra on spin magnitude $S$ is shown for $S=1/2$ to $S=\infty$ (LSW result) for $h=0.7568$ in Fig. \ref{Sgeq05}. Stronger $1/S$ corrections are observed along the $\Gamma$-$M$ line than the others, and more classical results are obtained for larger $S$. However, qualitatively the same behaviors are observed for various $S$. 
\subsection{Details of the roton spectrum}
\begin{figure*}[!t]
\begin{center}
\begin{minipage}{0.45\linewidth}
\begin{center}
\includegraphics[width=0.92\linewidth]{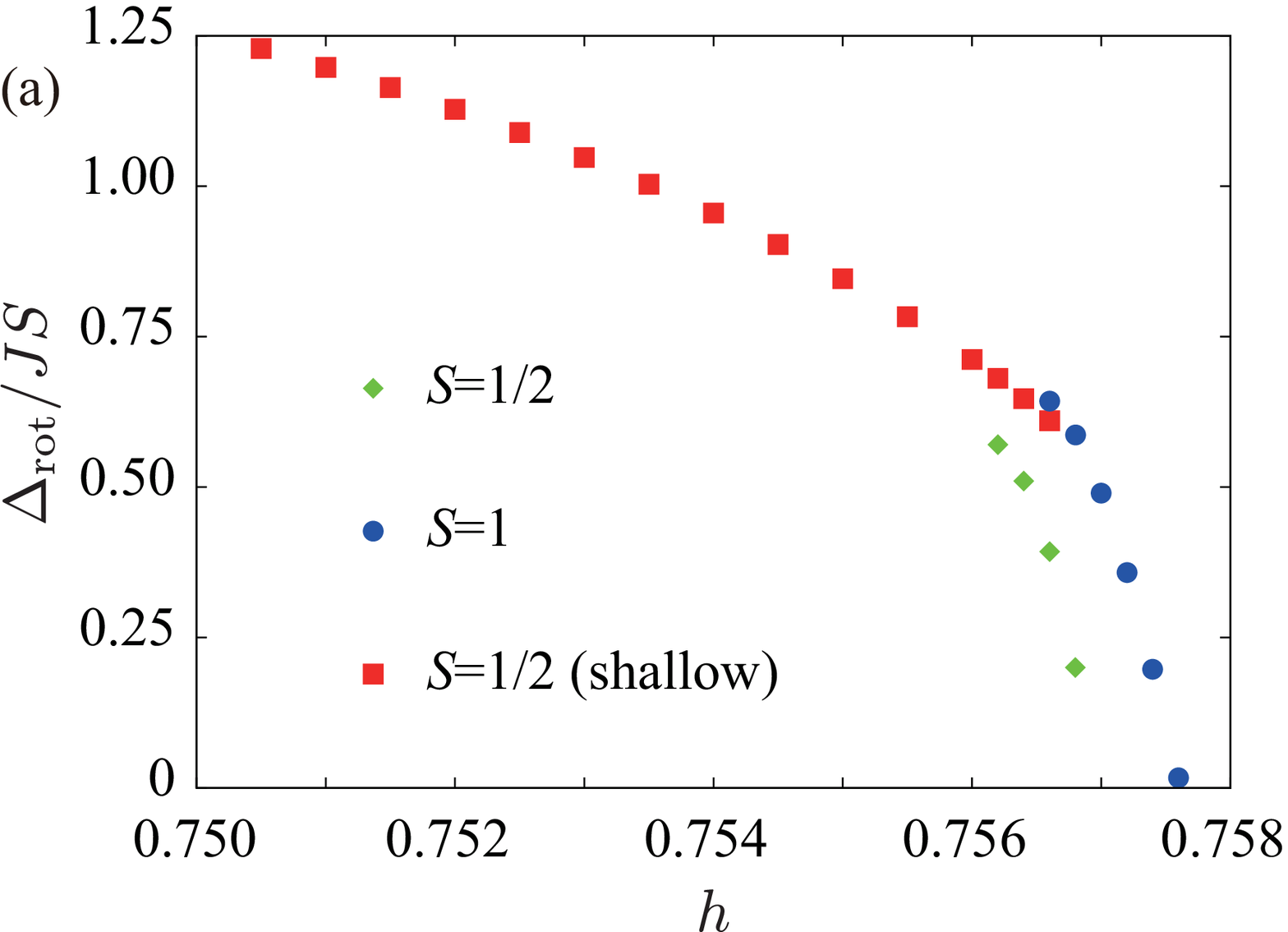}
\includegraphics[width=0.93\linewidth]{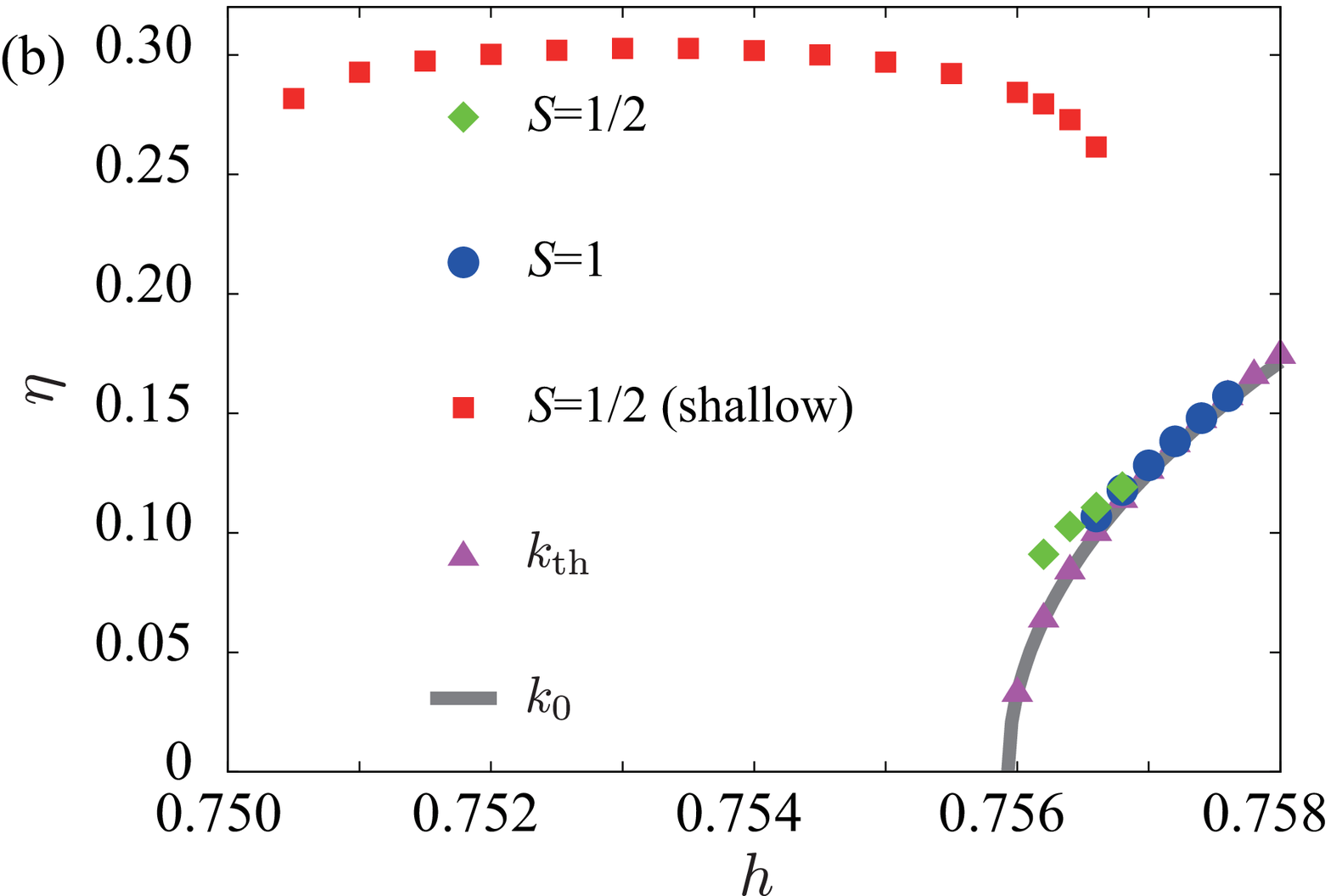}
\end{center}
\end{minipage}
\begin{minipage}{0.45\linewidth}
\begin{center}
\includegraphics[width=0.9\linewidth]{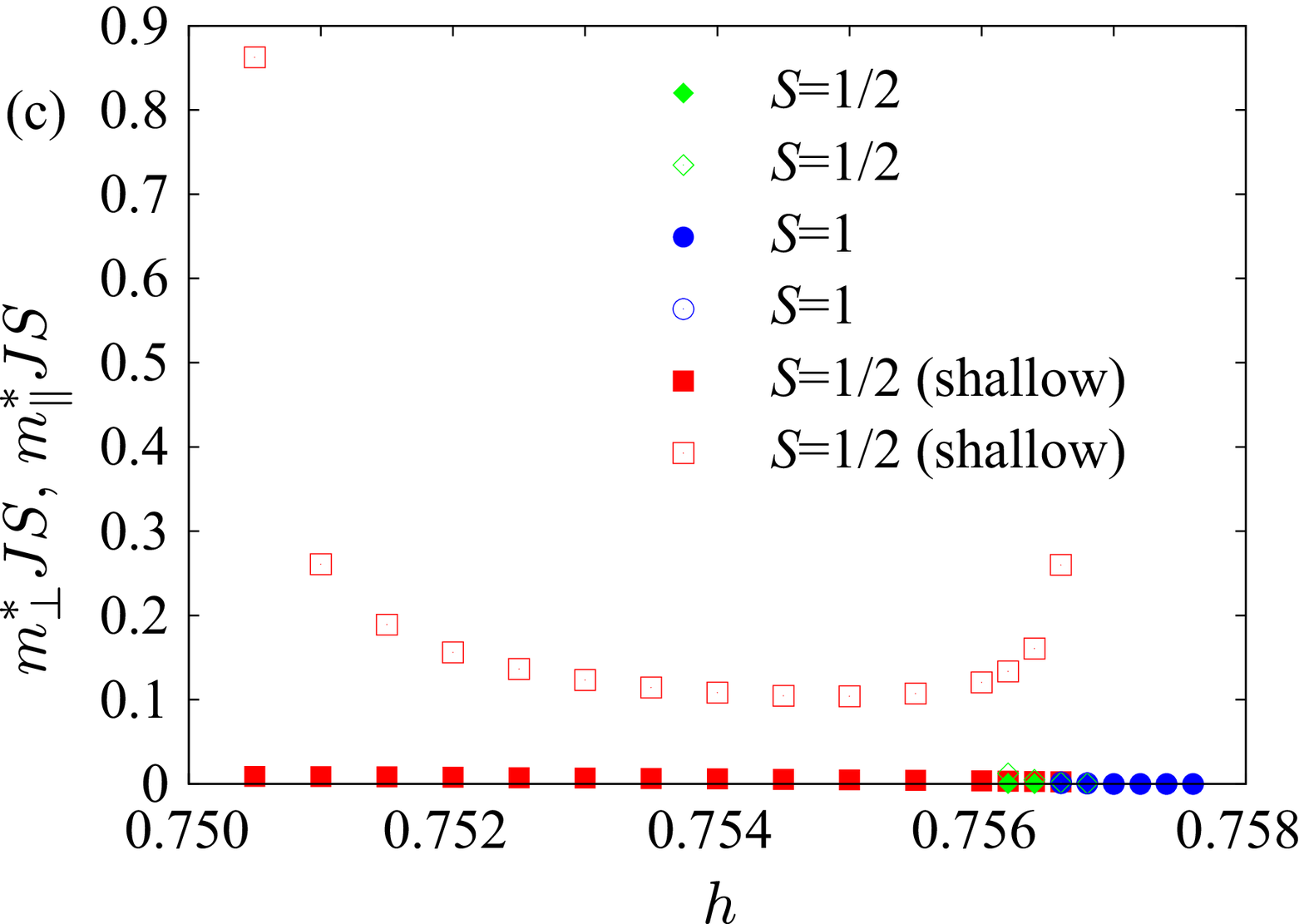}
\includegraphics[width=0.93\linewidth]{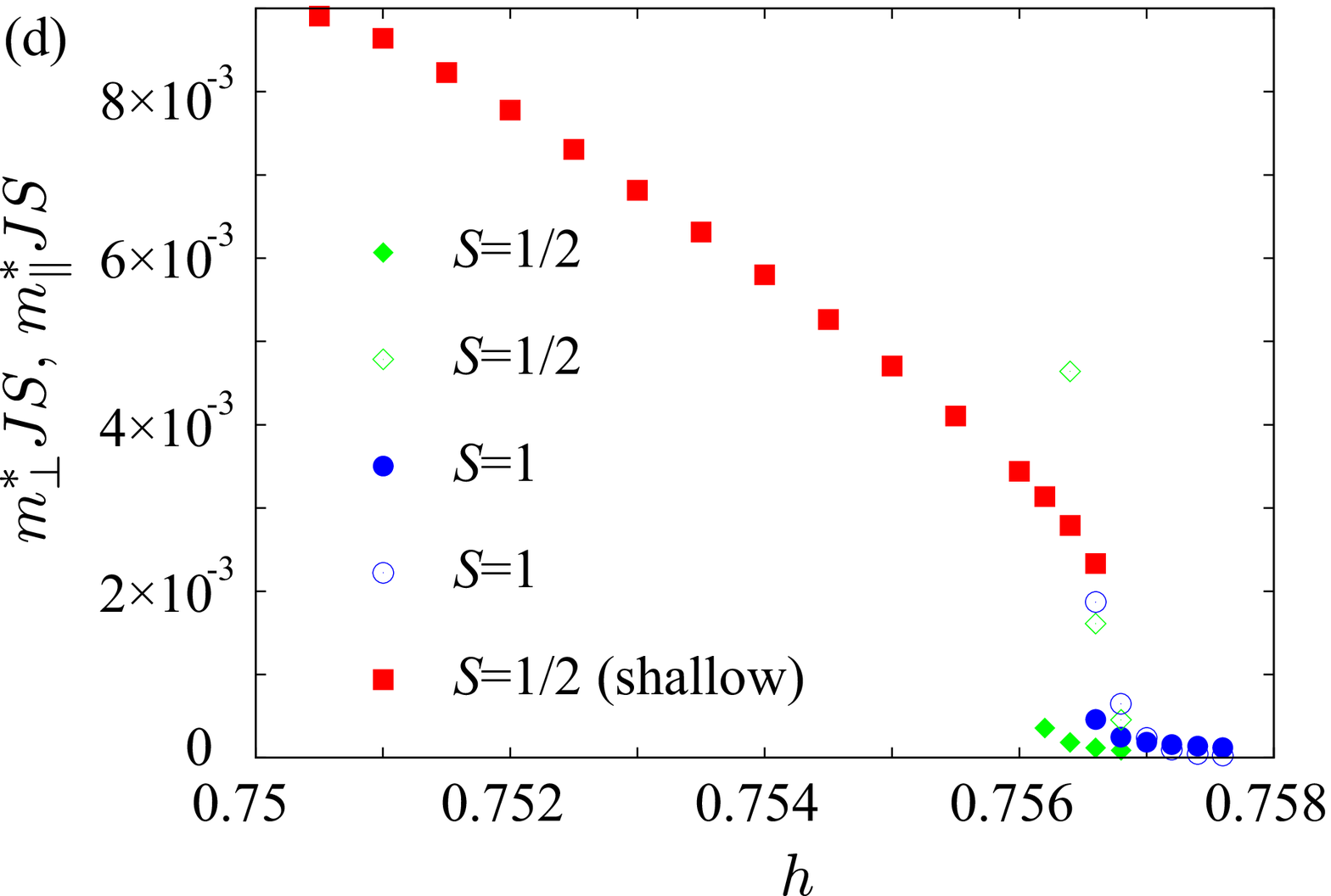}
\end{center}
\end{minipage}
\caption{(Color online) (a) Roton gap $\Delta_{\rm rot}$ along the $\Gamma$-$M$ line as a function of the field $h$. Solid diamonds, circles and squares represent results for $S=1/2$, $S=1$ rotons, and shallow rotons which appears only for $S=1/2$. Each roton gap decreases monotonically as the field increases. (b) Roton wave vector along the $\Gamma$-$M$ line [$\pi(1\!-\!\eta,1\!-\!\eta),0\leq \eta \leq1$] as a function of $h$. The triangles and gray line correspond to $k_{\rm rot}$ and $k_0$. We notice that $k_{\rm rot}$ appears in the neighborhood of $k_{\rm th}$. Little change in $k_{\rm rot}$ is observed for the shallow roton. (c) Roton masses as functions of $h$, where solid (open) points correspond to $m^{*}_{\perp}$ ($m^{*}_{\parallel}$). The masses are quite small, and nonmonotonic behavior is observed for the shallow roton. (d) Zoomed region of (c) is shown for clarity. The order of the ordinates is changed by two digits.}\label{k0kthkrot}
\end{center}
\end{figure*}

We examine how the roton, which appears only in the narrow range with a width less than 1\% of the saturation field, varies sensitively with field $h$. We determine the roton wavevector, gap, and mass by performing differentiations parallel and perpendicular to the $\Gamma$-$M$ line. Here, rotons emerge also for $S\geq3/2$, but we focus on $S=1/2$ and $S=1$ rotons for a while. The roton gap $\Delta_{\rm rot}$ and wave vector ${\bf \Tilde{k}}_{\rm rot}$ as functions of $h$ are shown in Figs. \ref{k0kthkrot}(a) and \ref{k0kthkrot}(b). We denote ${\bf q}$(${\bf \Tilde{q}}\!=\!{\bf Q}+{\bf q}$) as a wave vector measured from the $M$($\Gamma$) point from this section. The roton mass perpendicular (parallel) to the $\Gamma$-$M$ line $m^{*}_{\perp}$ ($m^{*}_{\parallel}$) as a function of $h$ is shown in Figs. \ref{k0kthkrot}(c) and \ref{k0kthkrot}(d), where solid points represent $m^{*}_{\perp}$ ( open points represent $m^{*}_{\parallel}$). 

We notice that $\Delta_{\rm rot}$ drops steeply to zero, reflecting the increasing three-magnon interactions. We also notice that ${\bf \Tilde{k}}_{\rm rot}$ appears approximately at the wave vector of the decay threshold ${\bf \Tilde{k}}_{\rm th}$, where the three-magnon interactions are especially strong because of the infinitesimal energy differences of the initial and intermediate states. Roton masses $m^{*}_{\perp}$ and $m^{*}_{\parallel}$ for both $S=1/2$ and $S=1$ also drop steeply, and $m^{*}_{\parallel}$ decrease more rapidly than $m^{*}_{\perp}$. It is also notable that roton masses are quite small, which means that an extremely sharp structure appears near ${\bf \Tilde{k}}_{\rm rot}$

In the case of the quantum limit, $S=1/2$, another roton (shallow roton) emerges, and its ${\bf \Tilde{k}}_{\rm rot}$ and $m^{*}_{\parallel}$ change nonmonotonically while $\Delta_{\rm rot}$ and $m^{*}_{\perp}$ decrease monotonically as $h$ increases. The former behavior is due to the competition between a smaller energy difference between the two states on the side of the $M$ point and a stronger spectral weight on the side of the $\Gamma$ point, and the latter is due to the increasing nonlinear interactions with increasing $h$. In addition, the shallow roton also emerges at around $h\approx0.5$ near the $\Gamma$ points along the $\Gamma$-$M$ line. However, the shallow rotons may easily disappear with finite interlayer couplings or finite temperature effects. Thus, we do not discuss them hereafter and focus on the ordinary rotons. 

First, we examine ${\bf \Tilde{k}}_{\rm rot}$ by using the fact that ${\bf \Tilde{k}}_{\rm rot}\approx{\bf \Tilde{k}}_{\rm th}$. The expanded LSW spectrum $\epsilon_{\bf \Tilde{k}}$ at around the $M$ point \cite{Mourigal,ZhitRMP} up to fifth order in $k$ along the $\Gamma$-$M$ line is given by
\begin{align}
\begin{split}
\epsilon_{{\bf \Tilde{k}}}&\approx ck(1+\alpha k^2+\beta k^4)\textrm{,}\\
c&=2JS\sqrt{2}\cos \theta \textrm{,}\\
\alpha&=\frac{1}{12\cos \theta}\left[\left(\frac{h}{h^{*}}\right)^2-1\right]\textrm{,}\\ 
\beta&=\frac{h^4 -32h^2+16}{7680\cos^2\theta}\textrm{.}\label{alphasgn}
\end{split}
\end{align}

The threshold wave vector ${\bf \Tilde{k}}_{\rm th}$ is determined by \cite{Mourigal}:
\begin{align}
\epsilon_{\bf \Tilde{k}}-2\epsilon_{\left({\bf \Tilde{k}}+{\bf Q}\right)/2}=0\textrm{.}\label{detk0}
\end{align}
Then, we approximate Eq. (\ref{detk0}) by using Eq. (\ref{alphasgn}):
\begin{align}
\epsilon_{\bf \Tilde{k}}-2\epsilon_{\left({\bf \Tilde{k}}+{\bf Q}\right)/2}\approx\frac{3}{4}k^3 c\left(\alpha+\frac{5}{4}\beta k^2 \right)\textrm{.}\label{decay}
\end{align}
We define an approximate value ${\bf \Tilde{k}}_0$ of the threshold wave vector ${\bf \Tilde{k}}_{\rm th}$ by setting the left-hand side of Eq. (\ref{decay}) to zero:
\begin{align}
k_0\!=\!\sqrt{\frac{-4\alpha}{5\beta}}\propto\left[\left(\frac{h}{h^{*}}\right)^2-1\right]^{1/2} (h^{*}\!\leq\! h\leq1)\textrm{.}\label{k0}
\end{align}
This approximation is valid for sufficiently small $k_0$. The wave vectors $k_{\rm th}$ and $k_0$ merge asymptotically for $h \rightarrow h^{*}$ in Fig. \ref{k0kthkrot}(b).

\begin{figure}[!h]
\begin{center}
\includegraphics[width=0.7\linewidth]{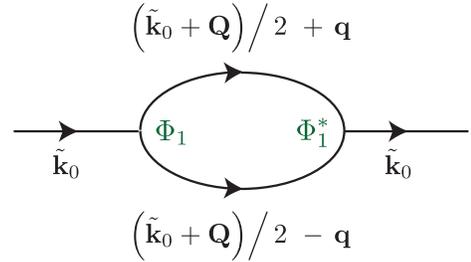}
\caption{(Color online) Self-energy causing strong renormalization near the roton wave vector.}\label{mixing}
\end{center}
\end{figure}

We now clarify the field dependence of $\Delta_{\rm rot}$, where a derivation is given in Appendix A. We focus on the lowest-order self-energy corrections shown in Fig. \ref{mixing}:
\begin{align}
\delta\epsilon_{{\bf \Tilde{k}}_0}=\Sigma^{(1)}({\bf \Tilde{k}}_0,\epsilon_{{\bf \Tilde{k}}_0})\label{DelEp}
\end{align}
in the limit of ${\bf q}\rightarrow0$, whose corrections are expected to be strong near ${\bf \Tilde{k}}_{\rm rot}$.

We perform the approximations
\begin{align}
\begin{split}
\sqrt{\frac{A_{\bf \Tilde{k}}+B_{\bf \Tilde{k}}}{A_{\bf \Tilde{k}}-B_{\bf \Tilde{k}}}}&=\sqrt{\frac{1-\gamma_{\bf k}}{1+\cos 2\theta \gamma_{\bf k}} } \approx \frac{k}{2\sqrt{2} \cos \theta} \textrm{,}\\
\gamma_{\bf \Tilde{k}} &\approx -1\textrm{,} \label{AppAkBk}
\end{split}
\end{align}
to the matrix elements of Eq. (\ref{DelEp}). We denote
\begin{align}
\Phi_1\!\left(\!{\bf \Tilde{k}_0},{\bf q}\!\right)\!=\!\Phi_1\!\left(\!{\bf \Tilde{k}}_0,\!\left(\!{\bf \Tilde{k}}_0\!+\!{\bf Q}\right)\!/2\!+\!{\bf q},\!\left({\bf \Tilde{k}}_0\!+\!{\bf Q}\right)\!/2\!-\!{\bf q}\!\right)\!\textrm{.}
\end{align}
for simplicity. We get
\begin{align}
\frac{\left.\left|\Phi_1\left({\bf \Tilde{k}}_0,{\bf q}\right)\right|^2\right|_{{\bf q}\rightarrow0}}{J^2S(\sin 2\theta)^2}
\approx \frac{k_0^3}{(2\sqrt{2}\cos \theta)^3}\textrm{.} \label{k03}
\end{align}

We then approximate the denominator of $\delta\epsilon_{{\bf \Tilde{k}}_0}$ \cite{Mourigal}: 
\begin{align}
\begin{split}
w[{\bf \Tilde{k}}_0,{\bf q}]&=\epsilon_{{\bf \Tilde{k}}_0}-\epsilon_{\left({\bf \Tilde{k}}_0+{\bf Q}\right)/2+{\bf q}}-\epsilon_{\left({\bf \Tilde{k}}_0+{\bf Q}\right)/2-{\bf q}} \\
&\approx -\frac{2c}{k_0}\frac{q^2}{1-(2q/k_0)^2}[\phi^2\!-\!\phi_0^2]\textrm{,}\\
\intertext{where $\phi$ denotes an azimuthal angle and}
\phi_0&=\sqrt{6\alpha}\frac{(k_0/2)^2-q^2}{q}\label{Wdef}\textrm{.}
\end{split}
\end{align}

We obtain lowest-order self-energy corrections:
\begin{align}
\begin{split}
\delta\epsilon_{{\bf \Tilde{k}}_0}&\propto -Jk_0^4 \tan^2 \theta \int \frac{q{\rm d}q}{q^2}\\
&= Jk_0^4 \tan^2 \theta \,\textrm{ln} \left[\frac{\Lambda}{k_0}\right]\textrm{,}\label{apgap}
\end{split}
\end{align} 
where $\Lambda$ denotes a lower cutoff. We see that the roton is related to a logarithmic factor, which originates from a two-dimensionality \cite{Mourigal,ZhitSCBApprox,Quasi2DZhitGroup,ZhitRMP}. 

\begin{figure}[!h]
\begin{center}
\begin{minipage}{0.9\linewidth}
\begin{center}
\includegraphics[width=0.87\linewidth]{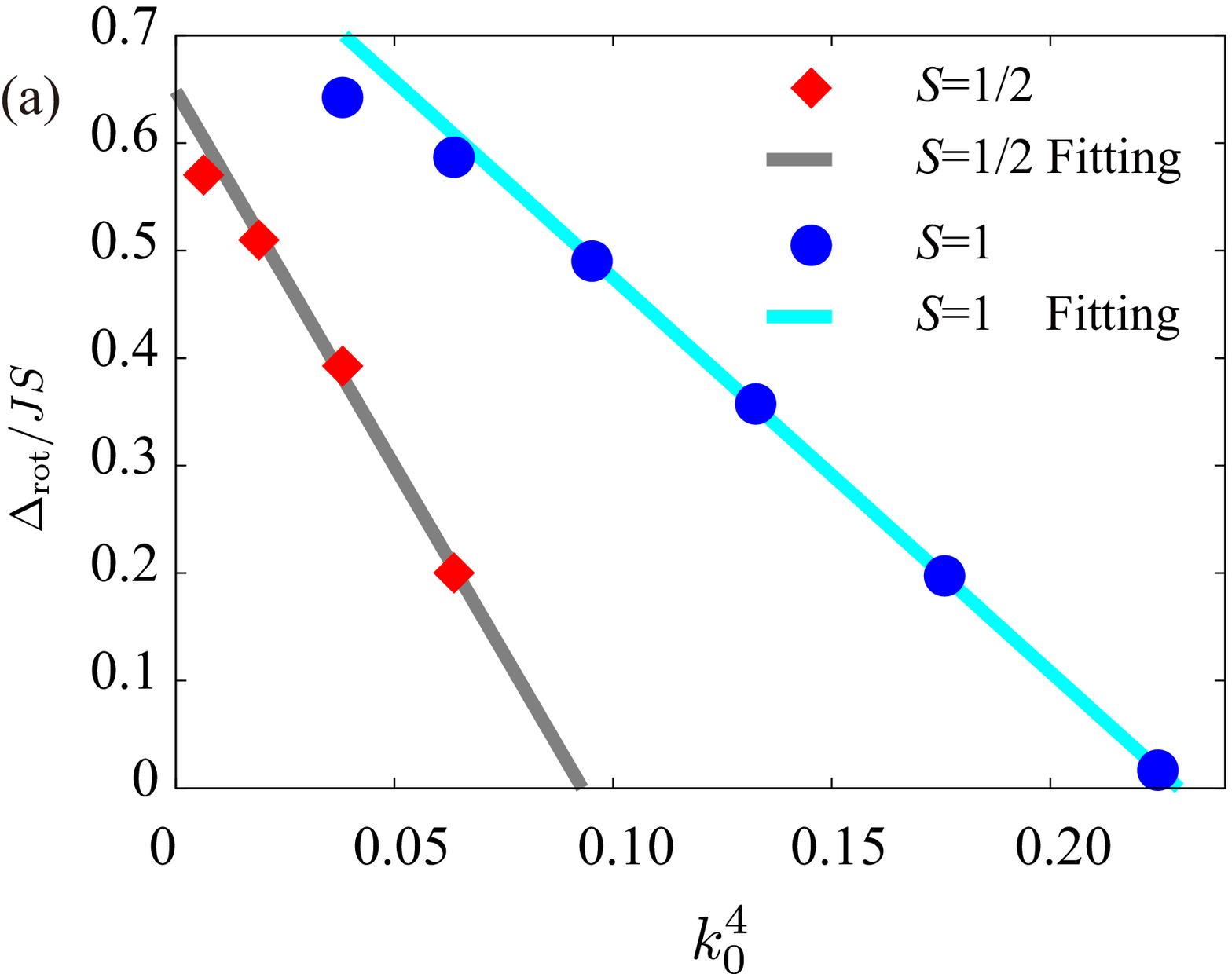}
\includegraphics[width=0.9\linewidth]{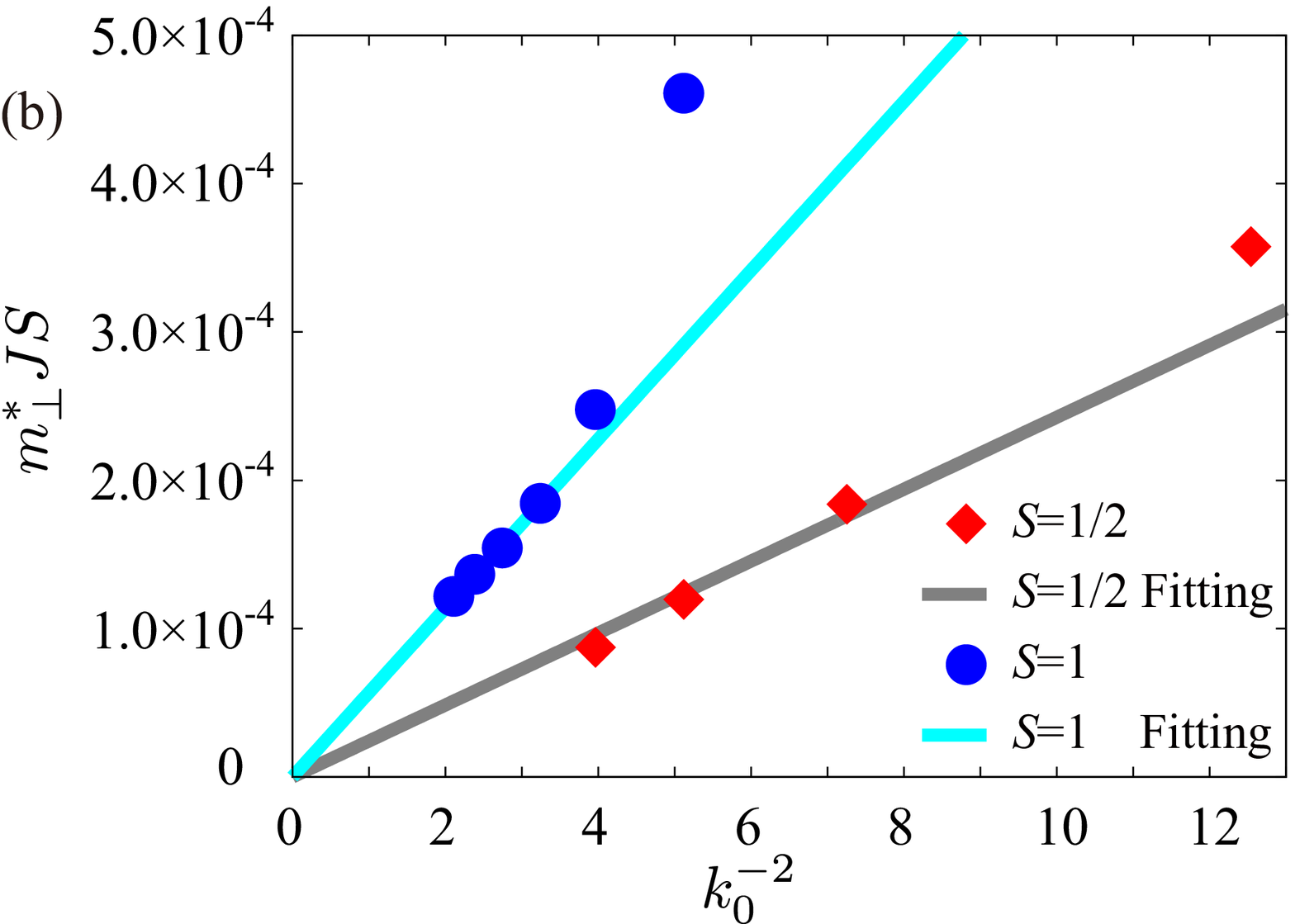}
\includegraphics[width=0.9\linewidth]{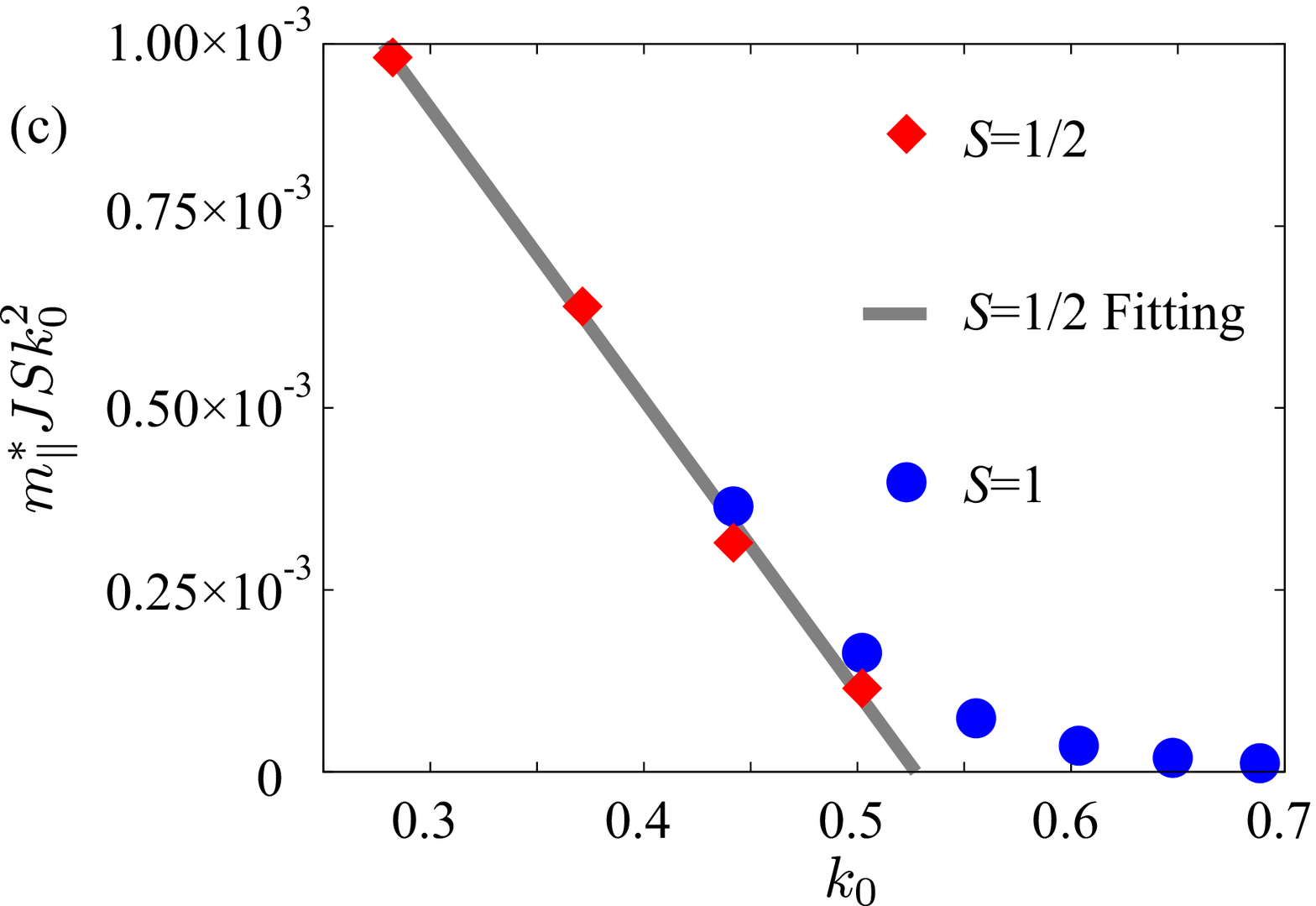}
\end{center}
\end{minipage}
\caption{(Color online) (a) Roton gap $\Delta_{\rm rot}$ as a function of $k_0^4$. Solid diamonds and circles represent $S=1/2$ and $S=1$, and lines show fittings using Eq. (\ref{ApproxGap}). The $S=1/2$ roton gap decreases about two times faster than the $S=1$ roton gap. (b) Perpendicular mass $m^{*}_{\perp}$ as a function of $k_0^{-2}$; it scales linearly with $k_0^{-2}$. (c) Parallel mass $m^{*}_{\parallel}$ plotted as $m^{*}_{\parallel}\,k^{2}_0$ vs $k_0$. We see that Eq. (\ref{MPalla}) gives a good approximation to $m^{*}_{\parallel}$ for sufficiently small $k_0$.}
\label{GapMas}
\end{center}
\end{figure}

Now, we get an approximation to $\Delta_{\rm rot}$:
\begin{align}
\Delta_{\rm rot} \approx ck_0 + A_0 \, Jk_0^4 \tan^2 \theta \textrm{ln} \left[\frac{\Lambda}{k_0}\right] \textrm{,}\label{ApproxGap}
\end{align}
where $A_0$ denotes a constant. Figure \ref{GapMas}(a) shows that $\Delta_{\rm rot}$ decreases proportionally to $k_0^4$. It is clear that the factor $k_0^4$ drives the softening. We also see that the logarithmic factor behaves almost as a constant in the region shown in Fig. \ref{GapMas}(a) and does not have any singularity there. 

The perturbation calculation still works well with a sufficiently small $k_0$. However, $k_0$ increases as $h/h^{*}$ increases [see Eq. (\ref{k0})], and finally, the perturbation calculation on the basis of the simple canted state is no longer valid (even for $S\geq3/2$), indicating a new ground state, which might be characterized by modulations with the roton wave vector ${\bf k}_{\rm rot}$. We thus see that the roton is essential in determining the new ground state at higher fields. In addition, the $S=1/2$ roton gap decreases about two times faster than the $S=1$ roton gap since the dependence on fields is attributed to $1/S$ corrections. 

We calculate the perpendicular effective mass $m^{*}_{\perp}$ by differentiating $\delta\epsilon_{\bf k}$:
\begin{align}
\begin{split}
m^{*}_{\perp} \propto k_0^{-2}\textrm{.}\label{mstar}
\end{split}
\end{align}
A derivation is given in Appendix B. The perpendicular mass $m^{*}_{\perp}$ is shown in Fig. \ref{GapMas}(b). It is clear that $m^{*}_{\perp}$ decreases monotonically with increasing $k_0$, and its behavior is well described by Eq. (\ref{mstar}). 

We also calculate the parallel mass $m^{*}_{\parallel}$: 
\begin{align}
m^{*}_{\parallel}\propto \left[k_0^2(1+b_0k_0)\right]^{-1}\approx k_0^{-2}(1-b_0k_0)\textrm{,}\label{MPalla}
\end{align}
where $b_0$ denotes a constant. This approximation is valid for sufficiently small $k_0$. The quantity $m^{*}_{\parallel}\,k_0^2$ is shown as a function of $k_0$ in Fig. \ref{GapMas}(c); it scales linearly with $k_0$ for sufficiently small $k_0$. 

We have calculated $\Delta_{\rm rot}$, $m^{*}_{\perp}$, and $m^{*}_{\parallel}$, and our approximations describe their field dependences rather well. In other words, the perturbation calculation and analytical approximation agree reasonably well.

We also discuss the valleylike structure perpendicular to the $\Gamma$-$M$ line in Fig. \ref{ContPlot}(c). Here, we examine $\delta\epsilon_{{\bf k}_0}$ given in Eq. (\ref{DelEp}). We approximate the energy denominator $w[{\bf \Tilde{k}}_0,{\bf q}]$ by using Eq. (\ref{Wdef}):
\begin{align}
\begin{split}
w[{\bf \Tilde{k}}_0,{\bf q}] \,\propto& -\frac{cqk^2_0}{2}\frac{\phi^2-\phi_0^2}{\phi_0}\textrm{.}
\end{split}
\end{align}
We evaluate the angular integration as
\begin{align}
\begin{split}
\int\! \frac{{\rm d}\phi}{w[{\bf \Tilde{k}}_0,{\bf q}] }\!&\approx\! -\frac{1}{cqk^2_0} \!\int\!{\rm d}\phi \left(\frac{1}{\phi\!-\!\phi_0}\!-\!\frac{1}{\phi\!+\!\phi_0}\right)\\
&\propto \, \textrm{ln}\left[\left|\phi-\phi_0\right| +\delta\phi \right]\textrm{,}
\end{split}
\end{align}
where $\delta\phi$ denotes a cutoff coming from the small difference between $k_{\rm rot}$ and $k_{\rm th}$:
\begin{align}
\delta\phi \!\propto\! k_{\rm rot}/k_{\rm th}-1\textrm{.} 
\end{align}
 
\section{DISCUSSION}
We have studied the excitation spectra of SLHAF in fields and have found the appearance and softening of the roton as a function of $h/h^{*}$. We consider that the roton emerges and softens as a precursor of the phase transition. Then, it seems interesting to determine the new ground state. In this section, the new ground state and possibilities to detect the roton are discussed. 
\subsection{New ground state}
In the classical limit $S=\infty$, the simple canted state is selected to minimize the exchange energy and the Zeeman energy. However, for finite $S$, the three-magnon interactions are induced by the noncollinear structures. Its effects are strong at around $h\approx0.75$, especially for the particular wave vector ${\bf \Tilde{k}}_{\rm rot}\approx{\bf \Tilde{k}}_{0}$ causing the roton's appearance. 

Previous works on $S=1/2$ SLHAFs calculate some static properties like the spin stiffness, magnetization, spin-wave velocity, and so on as functions of fields by exact diagonalization \cite{EDResult} and spin-wave calculation \cite{HighFieldLSW,Nikuni}, and no anomalies are observed at around $h\approx0.75$. However, a qualitative change between high and low fields is observed in the dynamical structure factor studied by quantum Monte Carlo simulation \cite{QMCResult} and exact diagonalization \cite{EDResult}. 

Considering these results, we expect that the new ground state might be rather similar to the simple canted states, and ${\bf k}_{\rm rot}$, where the transition occurs, may signify the new ground state. We speculate that a certain modulation with wave vector ${\bf k}_{\rm rot}$ occurs in the spin ordering in the $S_i^{x_0}$-$S_i^{y_0}$ plane while the canting angle $\theta$ remains essentially unchanged. In other words, the new ground state might be characterized by a freezing out of the roton mode. In addition, the new ground state might be similar to the spin-current order discussed in Ref. \onlinecite{Chubukov}. We believe that the roton's appearance and its softening are essentially correct, although they are not quantitatively accurate since these are the results of the second-order perturbation calculations.
\subsection{Possibilities of detecting rotons}
Now, we discuss the possibilities of confirming rotons by experiments and numerical calculations. In the same way as for rotons in helium \cite{RotonSH,HeRoton}, it is possible to detect the rotonlike minimum by inelastic neutron scattering along the $\Gamma$-$M$ line and specific-heat measurements under certain conditions. We need proper materials and high-accuracy measurements, as will be discussed below. 

We need materials which have a low enough $H_{\rm s}$ to achieve $h\approx0.75$. Here, the field range, where the roton emerges, is about 0.1\% of $H_{\rm s}$. Accordingly, we also need the uniformity and control of magnetic field with a precision of order $h\approx0.001$ at around $h\approx0.75$. 

We also need a very high accuracy in the wave vector to confirm the roton by experiments and numerical calculations. This is because of the very sharp structure of the roton spectrum, especially along the direction perpendicular to the $\Gamma$-$M$ line. If the resolution in the wave vector is worse than the sharpness of the roton, the excitation spectrum may acquire an apparent width because of the steepness of the roton spectrum. It should be noted that such a width is essentially independent of the magnon decay \cite{Mourigal,ZhitSCBApprox,ZhitRMP}. This may give at least a partial explanation of the anomalously large width reported in experiments\cite{Masuda} and numerical calculations\cite{EDResult,QMCResult}. 

We note it is also possible to detect rotons for $S\geq3/2$, but we need the higher resolution in wave vectors to detect them by neutron scattering due to sharper structures for larger spins. This is due to the small roton masses [see Eqs.(\ref{mstar}) and (\ref{MPalla})] and the growing imaginary part $\zeta_{\bf k}$ of the self-energy in higher fields \cite{Mourigal,ZhitSCBApprox}. Consequently, we see $S=1/2$, $1$ Heisenberg antiferromagnets are the best candidates for neutron scattering measurements. For specific-heat measurements, we do not have to worry about the small roton mass and $\zeta_{\bf k}$ for $S\geq3/2$. Therefore, $S\geq3/2$ materials are also good candidates for the specific-heat measurements. 

Recently, almost ideal $S=1/2$ two-dimensional SLHAFs have been synthesized \cite{PyzFamily,Tsyrulin}. They might be experimental candidates to examine the properties at high fields since these compounds have small magnetic anisotropies, interlayer couplings, and low enough saturation fields. 

We see that it is worth trying to detect the roton, which varies remarkably with slight changes of $h$, although there may be difficulties. Furthermore, the experiments may find what the new ground state looks like. In addition, it is also stimulating to detect the apparent minima at point $P$, which requires less uniformity and control of the field $h$.

It should be possible to confirm the roton also by numerical calculations. Quantum Monte Carlo simulations \cite{QMCResult} and exact diagonalization studies \cite{EDResult} have caught some anomaly which might be an indication of the sharp softening of the roton. We expect that a clearer structure of the roton or its softening can be confirmed by performing such calculations on sufficiently large lattices.
\section{CONCLUSION}
We have investigated the field dependence of the nonlinear spin-wave spectrum within the second-order perturbation calculation. We have found the rotonlike minimum in the renormalized spectrum at the roton wave vector $k_{\rm rot}\approx k_0$, where three-magnon couplings are particularly strong. We have also calculated $\Delta_{\rm rot}$ and $m^{*}_{\perp}$ and found that they change as functions of $k_0$ (or $h/h^{*}$). 

We see that the especially strong three-magnon coupling near the point where the decay threshold meets the $\Gamma$-$M$ line causes the appearance of rotons accompanied by the logarithmic factor. Furthermore, the coupling increases as the field increases, triggering the softening of the roton. Thus, we consider that the roton is physically quite important. Although deciding the modulated ground state is beyond the scope of this paper, we expect that ${\bf k}_{\rm rot}$, when the modulation occurs, signifies what the new ground state is like. 

We have also found the valleylike structure perpendicular to the $\Gamma$-$M$ line near the ${\bf k}_{\rm rot}$. We see that the sharp structure near ${\bf k}_{\rm rot}$ may induce the apparent linewidth of experiments and numerical results \cite{EDResult,QMCResult,Masuda}. We expect that the anomalously large linewidth reported in previous works \cite{EDResult,QMCResult,Masuda} might be partially explained by its sharp structure even in the absence of magnon decay. 

The emergence and the softening the roton feature in the spin-wave spectrum is quite important at low-temperatures and high fields. By carefully tuning the magnetic field, it will be possible to see the roton effects in low temperature specific heats, neutron scattering, and spin transport \cite{Kubo}. Among other things we expect that the rotons may be most easily detected as an exponential temperature dependence in thermal and transport properties. Last, we expect that the roton feature can also be confirmed by careful numerical calculations on sufficiently large lattices.  
\section*{Acknowledgment}
This work was supported in part by Grants for Excellent Graduate Schools, MEXT, Japan.
\appendix
\section{ESTIMATION OF ROTON GAP}
We derive how $\Delta_{\rm rot}$ depends on $k_0$. We perform the approximation given in Eq. (\ref{AppAkBk}) to the matrix elements of $\delta\epsilon_{\bf \Tilde{k}_0}$ obtaining
\begin{align}
\begin{split}
\frac{2NS\left|\Phi_1({\bf \Tilde{k}},{\bf \Tilde{q}},{\bf \Tilde{p}})\right|^2}{(H \cos \theta)^2} \approx& \frac{9\,\Tilde{k}\Tilde{q}\Tilde{p}}{4\,(2\sqrt{2}\cos \theta)^3}
\textrm{,}\label{Num}
\end{split}
\end{align}
where $N$ denotes the number of sites and ${\bf \Tilde{p}}\!={\bf Q}+\!{\bf \Tilde{k}}\!-\!{\bf \Tilde{q}}$. 

Then, we consider the process in Fig. \ref{mixing}. We write $\delta\epsilon_{\bf k}$ as
\begin{align}
\begin{split}
\delta\epsilon_{{\bf \Tilde{k}}_0}&\approx \frac{1}{2}\sum_{\bf q}\frac{\left|\Phi_1({\bf \Tilde{k}}_0,{\bf q})\right|^2}{w[{\bf \Tilde{k}}_0,{\bf q}]}\label{k2plmi}
\end{split}
\end{align}
for simplicity. Then we consider the numerator in the limit of ${\bf q}\rightarrow 0$ using Eqs. (\ref{Num}) and (\ref{k2plmi}):
\begin{align}
\begin{split}
\frac{2NS\left|\Phi_1({\bf \Tilde{k}}_0,{\bf q})\right|^2|_{{\bf q}\rightarrow 0}}{(H \cos \theta)^2}
\approx&
\frac{9}{16}\frac{k_0^3}{(2\sqrt{2}\cos \theta)^3}\textrm{,}\label{element2q0}
\end{split}
\end{align}
and we obtain Eq. (\ref{k03}).

The denominator $w[{\bf \Tilde{k}}_0,{\bf q}]$ is approximated by \cite{Mourigal}
\begin{align}
|{\bf k}/2+{\bf q}|=k/2+q-\frac{1}{4}\frac{kq}{(k/2+q)^2}\phi^2\textrm{.}
\end{align}
We get Eq. (\ref{Wdef}) by using:
\begin{align}
\begin{split}
w[{\bf \Tilde{k}}_0,{\bf q}] \!\approx&c\left[k_0\!-\!\left|{\bf k}_0/2\!+\!{\bf q}\right|\!-\left|{\bf k}_0/2\!-\!{\bf q}\right|\right.\\
&\!+\!\alpha \!\left.\left(k_0^3\!-\!\left|{\bf k}_0/2\!+\!{\bf q}\right|^3\!-\!\left|{\bf k}_0/2\!-\!{\bf q}\right|^3\right)\right]\label{appep}\textrm{.}
\end{split}
\end{align}
We now obtain Eq. (\ref{apgap}) by using Eqs. (\ref{k03}) and (\ref{Wdef}).

\section{ESTIMATION OF ROTON MASS}
We obtain the $k_0$ dependence of $m_{\perp}$($m_{\parallel}$) by differentiating $\delta\epsilon_{\bf k}$ perpendicular(parallel) to the $\Gamma$-$M$ line.

We differentiate $\bar{\epsilon}_{\bf k}$ perpendicular to the line,
\begin{align}
\begin{split}
1/m^{*}_{\perp}\approx&\frac{1}{k_0}\frac{\partial \delta\epsilon_{\bf \Tilde{k}_0}}{\partial k_0}\\
=&\frac{1}{2k_0}\!\sum_{\bf q}\left[\!\frac{\partial}{\partial k_0}\!\left(\!\frac{1}{w[{\bf \Tilde{k}}_0,{\bf q}]}\right) \!\left|\Phi_1({\bf \Tilde{k}}_0,{\bf q})\right|^2\right.\\
&\hspace{10mm}\!+\left.\left(\frac{1}{ w[{\bf \Tilde{k}}_0,{\bf q}]}\right)\! \frac{\partial \left|\Phi_1({\bf \Tilde{k}}_0,{\bf q})\right|^2}{\partial k_0}\right]\!\textrm{,}\label{DifPer}
\end{split}
\end{align}
and parallel to the line,
\begin{align}
\begin{split}
1/m^{*}_{\parallel}\approx&\,\frac{\partial^2 (\epsilon_{{\bf \Tilde{k}}_0}+\delta\epsilon_{{\bf \Tilde{k}}_0})}{\partial k_0^2}\\
=&\,d_0k_0^3\\
&\!+\!\frac{1}{2}\!\sum_{\bf q}\!\left[\!\frac{\partial}{\partial k_0}\!\left(\!\frac{2}{ w[{\bf \Tilde{k}}_0,{\bf q}]}\right)\! \frac{\partial\! \left|\Phi_1\!({\bf \Tilde{k}}_0,{\bf q})\right|^2}{\partial k_0}\right.\\
&\hspace{9mm}+\!\frac{1}{w[{\bf \Tilde{k}}_0,{\bf q}]} \frac{\partial^2 \left|\Phi_1({\bf \Tilde{k}}_0,{\bf q})\right|^2}{\partial k_0^2}\\
&\hspace{9mm}+\!\left.\left|\Phi_1({\bf \Tilde{k}}_0,{\bf q})\right|^2\!\frac{\partial^2}{\partial k^2_0}\!\left(\frac{1}{w[{\bf \Tilde{k}}_0,{\bf q}]}\right)\!\right]\!\textrm{,}\label{Dif}
\end{split}
\end{align}
where $d_0$ denotes a constant, and we use 
\begin{align}
\begin{split}
\frac{\partial^2 \epsilon_{{\bf k}_0}}{\partial k^2_0}\approx&\,\frac{\partial^2}{\partial k^2_0}\left(ck_0+\frac{\alpha k_0^3}{5}\right)\\
\propto&\,k_0^3\textrm{.}
\end{split}
\end{align}
We obtain the following derivatives of $1/w[\Tilde{\bf k}_0,{\bf q}]$:
\begin{align}
\begin{split}
\frac{\partial}{\partial k_0}\left(\frac{1}{w[{\bf \Tilde{k}}_0,{\bf q}]}\right)=&-\frac{1}{w[{\bf \Tilde{k}}_0,{\bf q}]^2}\frac{\partial w[{\bf \Tilde{k}}_0,{\bf q}]}{\partial k_0}\textrm{,}\\
\frac{\partial^2}{\partial^2 k_0}\left(\frac{1}{w[{\bf \Tilde{k}}_0,{\bf q}]}\right)=&\frac{2}{w[{\bf \Tilde{k}}_0,{\bf q}]^3}\left(\frac{\partial w[{\bf \Tilde{k}}_0,{\bf q}]}{\partial k_0}\right)^2\\
&-\frac{1}{w[{\bf \Tilde{k}}_0,{\bf q}]^2}\frac{\partial^2 w[{\bf \Tilde{k}}_0,{\bf q}]}{\partial k^2_0}\textrm{.}\label{Dif2}
\end{split}
\end{align}
and using Eq. (\ref{Wdef}),
\begin{align}
\begin{split}
\frac{\partial w[{\bf \Tilde{k}}_0,{\bf q}]}{\partial k_0}\!\approx& \frac{2c q^2}{k_0^2}(\phi^2+3\phi_0^2)\textrm{,}\\
\frac{\partial^2 w[{\bf \Tilde{k}}_0,{\bf q}]}{\partial k_0^2}\!\approx\!&-\frac{4c q^2}{k_0^3}(\phi^2-3\phi_0^2)\textrm{.}\label{element1}
\end{split}
\end{align}
We now obtain $m_{\perp}$ and $m_{\parallel}$ by using Eqs. (\ref{DifPer}), (\ref{Dif}), (\ref{Dif2}), (\ref{element1}), and derivatives of matrix elements [see Eq. (\ref{k03})] with respect to $k_0$. We now obtain Eqs. (\ref{mstar}) and (\ref{MPalla}). 

\bibliography{refPRB}

\end{document}